\documentclass[prl,amsmath,amssymb,twocolumn]{revtex4}

\input epsf.tex
\usepackage{graphicx}

\newcommand{\be}{\begin{equation}}
\newcommand{\ee}{\end{equation}}
\newcommand{\ba}{\begin{eqnarray}}
\newcommand{\ea}{\end{eqnarray}}
\newcommand{\ban}{\begin{eqnarray*}}
\newcommand{\ean}{\end{eqnarray*}}

\newcommand{\sandwich}[3]{\mbox{$ \langle #1 | #2 | #3 \rangle $}}
\newcommand{\ket}[1]{\mbox{$ | #1 \rangle $}}

\newcommand{\one}{\leavevmode\hbox{\small1\normalsize\kern-.33em1}}

\begin{document}

\title{Experimental Falsification of Leggett's Non-Local Variable Model}
\author{Cyril Branciard$^1$, Alexander Ling$^{2,3}$, Nicolas Gisin$^1$, Christian Kurtsiefer$^2$, Antia Lamas-Linares$^2$, Valerio Scarani$^2$}
\affiliation{$^1$ Group of Applied Physics, University of Geneva, Switzerland\\
$^2$ Centre for Quantum Technologies, National University of
Singapore, Singapore\\ $^3$ Temasek Laboratories, National University
  of Singapore, Singapore}
\date{\today}
\begin{abstract}
Bell's theorem guarantees that no model based on local variables can
reproduce quantum correlations. Also some models based on
\textit{non-local variables}, if subject to apparently
``reasonable'' constraints, may fail to reproduce quantum physics.
In this paper, we introduce a family of inequalities, which allow
testing Leggett's non-local model versus quantum physics, and which
can be tested in an experiment without additional assumptions. Our
experimental data falsify Leggett's model and are in agreement with
quantum predictions.
\end{abstract}
\maketitle

\textit{Introduction.} Quantum physics provides a precise rule to
compute the probability that the measurement of $A$ and $B$
performed on two physical systems in the state $\ket{\Psi}$ will
lead to the outcomes $(r_A,r_B)$: \ba
P_Q(r_A,r_B|A,B)&=&\sandwich{\Psi}{{\cal P}_{r_A}\otimes {\cal
P}_{r_B} }{\Psi} \ea where ${\cal P}_{r}$ is the projector on the
subspace associated to the measurement result $r$. For entangled
states, this formula predicts that the outcomes are correlated,
irrespective of the distance between the two measurement devices. A
natural explanation for correlations established at a distance is
pre-established agreement: the two particles have left the source
with some common information $\lambda$, called a local variable
(LV), that allows them to compute the outcomes for each possible
measurement; formally, $r_A=f_A(A,\lambda)$ and
$r_B=f_B(B,\lambda)$. Satisfactory as it may seem a priori, this
model fails to reproduce all quantum correlations: this is the
celebrated result of John Bell \cite{bell}, by now tested in a very
large number of experiments. The fact that quantum correlations can
be attributed neither to LV nor to communication below the speed of
light is referred to as \textit{quantum non-locality}.

While non-locality is a striking manifestation of quantum
entanglement, it is not yet clear how fundamental this notion really
is: the essence of quantum physics may be somewhere else
\cite{pr94}. For instance, non-determinism is another important
feature of quantum physics, with no a priori link with non-locality.
Generic theories featuring both non-determinism and non-locality
have been studied, with several interesting achievements
\cite{nlnd}; but it is not yet clear what singles quantum physics
out. In order to progress in this direction, it is important to
learn which other alternative models are compatible with quantum
physics, which are not. Bell's theorem having ruled out all possible
LV models, we have to move on to models based on \textit{non-local
variables} (NLV). The first example of testable NLV model was the
one by Suarez and Scarani \cite{sua97}, falsified in a series of
experiments a few years ago \cite{zbi01}. A different such model was
proposed more recently by Leggett \cite{leg03}. This model supposes
that the source emits product quantum states
$\ket{\alpha}\otimes\ket{\beta}$ with probability density
$\rho(\alpha,\beta)$, and enforces that the marginal probabilities
must be compatible with such states: \ba
P(r_A|A)&=&\int d\rho(\alpha,\beta)\sandwich{\alpha}{{\cal P}_{r_A}}{\alpha}\,,\label{assa}\\
P(r_B|B)&=&\int d\rho(\alpha,\beta)\sandwich{\beta}{{\cal
P}_{r_B}}{\beta}\,.\label{assb} \ea The correlations however must
include some non-local effect, otherwise this would be a
(non-deterministic) LV model and would already be ruled out by
Bell's theorem. What Leggett showed is that the simple requirement
of consistency (i.e., no negative probabilities should appear at any
stage) constrains the possible correlations, even non-local ones, to
satisfy inequalities that are slightly but clearly violated by
quantum physics. A recent experiment \cite{gro07} demonstrated that
state-of-the-art setups can detect this violation in principle.
However, their falsification of the Leggett model is flawed by the
need for additional assumptions, because the inequality they used
\cite{suppl}, just as the original one by Leggett, supposes that
data are collected from infinitely many measurement settings. In
this paper, we present a family of inequalities, which allow testing
Leggett's model against quantum physics with a finite number of
measurements. We show their experimental violation by pairs of
polarization-entangled photons. We conclude with an overview of what
has been learned and what is still to be learned about NLV models.

\textit{Theory.} We restrict our theory to the case where the
quantum degree of freedom under study is a qubit. We consider von
Neumann measurements, that can be labeled by unit vectors in the
Poincar\'e sphere $\mathcal{S}$: $A\rightarrow \vec{a}$ and
$B\rightarrow \vec{b}$; their outcomes will be written
$r_A,r_B\in\{+1,-1\}$. Pure states of single particles can also be
labeled by unit vectors $\vec{u},\vec{v}$ in $\mathcal{S}$.
Leggett's model requires \footnote{The specific form of the marginal
distributions is called \textit{Malus' law} in the case of
polarization.} \ba P(r_A,r_B|\vec{a},\vec{b})&=&\int
d\rho(\vec{u},\vec{v})P_{\vec{u},\vec{v}}(r_A,r_B|\vec{a},\vec{b})\ea
with \ba
P_{\vec{u},\vec{v}}(r_A,r_B|\vec{a},\vec{b})&=&\frac{1}{4}\Big[1+r_A\vec{a}\cdot\vec{u}+r_B\vec{b}\cdot\vec{v}\nonumber\\&&+r_Ar_B
C(\vec{u},\vec{v},\vec{a},\vec{b})\Big]\,.\label{Puv} \ea The
correlation coefficient $C(\vec{u},\vec{v},\vec{a},\vec{b})$ is
constrained only by the requirement that (\ref{Puv}) must define a
probability distribution over $(r_A,r_B)$ for all choice of the
measurements $\vec{a},\vec{b}$. Remarkably, this constraint is
sufficient to derive inequalities that can be violated by quantum
physics \cite{leg03}. The inequality derived in \cite{suppl} (see
also \cite{par07} for a subsequent shorter derivation) reads \ba
\left|E_1(\varphi)+E_1(0)\right|+\left|E_2(\varphi)+E_2(0)\right|&\leq
& 4-\frac{4}{\pi}\left|\sin\frac{\varphi}{2}\right|
\label{ineqorig}\ea where the quantities $E_j(\theta)$ are defined
from the correlation coefficients \ba
C(\vec{a},\vec{b})&=&\sum_{r_A,r_B}r_Ar_B P(r_A,r_B|\vec{a},\vec{b})
\ea as follows. The index $j$ refers to a plane $\{\vec{a} \in
\mathcal{S} | \vec{a} \cdot \vec{n}_j\ = 0\}$ in the Poincar\'e
sphere (for $\vec{n}_j \in \mathcal{S}$),
and the two planes $j=1,2$ that appear in (\ref{ineqorig}) must be
orthogonal (i.e. $\vec{n}_1 \cdot \vec{n}_2 = 0$).
For each unit vector $\vec{a}_j$ of plane $j$, let's define
$\vec{a}_j^{\,\perp} = \vec{n}_j \times \vec{a}_j$. $E_j(\theta)$ is
then the average of $C(\vec{a}_j,\vec{b}_j)$ over \textit{all}
directions $\vec{a}_j$, with
$\vec{b}_j=\cos\theta\,\vec{a}_j+\sin\theta\,\vec{a}_j^{\,\perp}$
\footnote{This step is taken after (27) in \cite{suppl}, before (8)
in \cite{par07}. The derivation of the original inequalities goes
through the same step between (3.9) and (3.10) in \cite{leg03}.}.
This is a problematic feature of inequality (\ref{ineqorig}): it can
be checked only by performing an infinite number of measurements or
by adding the assumption of rotational invariance of the correlation
coefficients $C(\vec{a},\vec{b})$, as in \cite{gro07}. It is thus
natural to try and replace the average over all possible settings
with an average on a discrete set. This is done by the following
estimate. Let $\vec{w}$ and $\vec{c}$ be two unit vectors, and let
$R_N$ be the rotation by $\frac{\pi}{N}$ around the axis orthogonal
to ($\vec{w}$,$\vec{c}$). Then \ba
\frac{1}{N}\sum_{k=0}^{N-1}\left|\left(R_N^k\,\vec{c}\right)\cdot\vec{w}\right|&\geq
& u_N =\frac{1}{N}\cot\frac{\pi}{2N}\,.\label{discr} \ea Indeed, let
$\tilde{\xi}$ be the angle between $\vec{w}$ and $\vec{c}$, and $\xi
= (\tilde{\xi} - \frac{\pi}{2}) \bmod \frac{\pi}{N}$, such that $\xi
\in [0,\frac{\pi}{N}[$: then it holds $\sum_{k = 0}^{N-1} |(R_N^k
\vec{c}) \cdot \vec{w} | = \sum_{k = 0}^{N-1} |\cos(\tilde{\xi} +
\frac{k\pi}{N}) | = \sum_{k = 0}^{N-1} \sin(\xi + \frac{k\pi}{N}) =
\sin \xi + Nu_N \cos \xi  \geq Nu_N$ as announced.

Replacing the full average by the discrete average (\ref{discr}) in
the otherwise unchanged proofs \cite{suppl,par07}, we obtain the
following family of inequalities: \ba
\left|E_1^N(\vec{a}_1,\varphi)+E_1^N(\vec{a}_1,0)\right|&+&\left|E_2^N(\vec{a}_2,\varphi)+E_2^N(\vec{a}_2,0)\right|\nonumber\\\equiv
L_N(\vec{a}_1,\vec{a}_2,\varphi)&\leq &
4-2u_N\left|\sin\frac{\varphi}{2}\right| \label{ineqdiscr}\ea where
\ba
E_j^N(\vec{a}_j,\theta)&=&\frac{1}{N}\sum_{k=0}^{N-1}C\big(\vec{a}_j^{\,k},\vec{b}_j^{\,k}\big)\label{EjN}
\ea with
$\vec{b}_j=\cos\theta\,\vec{a}_j+\sin\theta\,\vec{a}_j^{\,\perp}$
and the notation $\vec{c}^{\,k}=(R_{N,j})^k\,\vec{c}$ (the
$\frac{\pi}{N}$-rotation is along $\vec{n}_j$). This defines $2N$
and $4N$ settings on each side. For a pure singlet state, the
quantum mechanical prediction for $L_N(\vec{a}_1,\vec{a}_2,\varphi)$
is \ba L_{\Psi^-}(\varphi)&=&2(1+\cos\varphi)\label{singlet} \ea
independent of $N$ and of the choice of $\vec{a}_1,\vec{a}_2$ since
the state is rotationally invariant.

\begin{figure}
\centerline{\epsffile{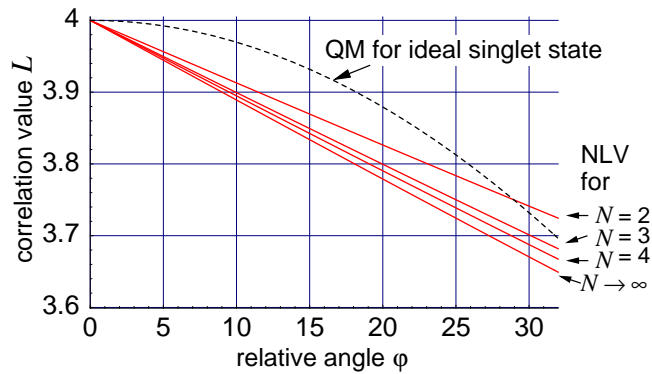}}
\caption{\label{fig:theoimage} Dependency of the combined
correlation parameters
  $L(\varphi)$ as a function of the separation angle $\varphi$ for the quantum mechanical prediction for a
  pure singlet state, and bounds for non-local
 variable models assuming an averaging over various numbers of
 directions $N$.
}
\end{figure}

The inequality for $N=1$ cannot be violated because $u_1=0$
\footnote{Actually, the data measured on a singlet state for $N=1$,
as in \cite{gro07}, can be reproduced by the explicit NLV
Leggett-type model presented in \cite{suppl}. Indeed, the validity
condition for that NLV model is that there exists unit vectors
$\vec{u}, \vec{v}$ in the Poincar\'e sphere such that, for all pairs
of observables $\vec{a}, \vec{b}$ measured in the experiment, one
has $|\vec{a} \cdot \vec{b} \pm \vec{u} \cdot \vec{a}| \leq 1 \mp
\vec{v} \cdot \vec{b}$ (Eq.~(10) of \cite{suppl}) or, equivalently,
$|\vec{a} \cdot \vec{b} \pm \vec{v} \cdot \vec{b}| \leq 1 \mp
\vec{u} \cdot \vec{a}$. Now, for the case $N = 1$, one would measure
four sets of observables $\vec{a}_j$, $\vec{b}_j = \cos \theta
\vec{a_j} + \sin \theta \vec{a_j}^\perp$ in planes $j = 1,2$ and for
$\theta = 0, \varphi$. Then for $\vec{u} = -\vec{v}$ orthogonal to
both $\vec{a}_1^\perp$ and $\vec{a}_2^\perp$ and whatever $\theta$,
one has $|\vec{a_j} \cdot \vec{b_j} \pm \vec{v} \cdot \vec{b_j}| =
|\cos \theta \mp \vec{u} \cdot (\cos \theta \vec{a_j} + \sin \theta
\vec{a_j}^\perp)| = |\cos \theta (1 \mp \vec{u} \cdot \vec{a_j})|
\leq 1 \mp \vec{u} \cdot \vec{a_j}$ as required.}. Already for
$N=2$, however, quantum physics violates the inequality: this opens
the possibility for our falsification of Leggett's model without
additional assumptions \footnote{Note that, since the model under
test is NLV, there are no such concerns as locality or memory
loopholes. The detection loophole is obviously still open.}. For
$N\rightarrow\infty$, $u_N\rightarrow \frac{2}{\pi}$: one recovers
inequality (\ref{ineqorig}). The suitable range of difference angles
$\varphi$ for probing a violation of the inequalities
(\ref{ineqdiscr}) can be identified from figure~\ref{fig:theoimage}.
The largest violation for an ideal singlet state would occur for
$|\sin \frac{\varphi}{2}| = \frac{u_N}{4}$, i.e. at
$\varphi=14.4^\circ$ for $N=2$, increasing with $N$ up to
$\varphi=18.3^\circ$ for $N\rightarrow\infty$.

\begin{figure}
\centerline{\epsffile{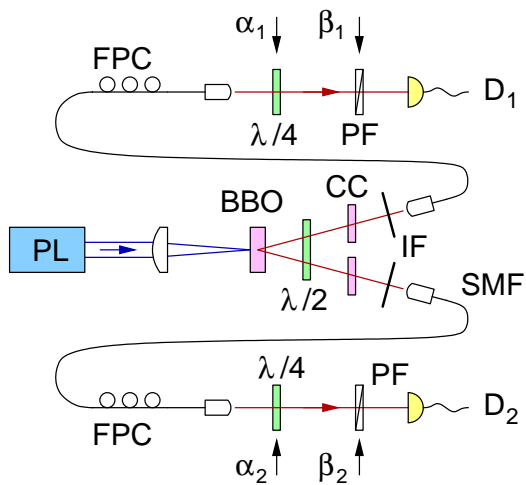}}
\caption{\label{expsetup}Experimental setup. Polarization-entangled photon
  pairs are generated in Barium-beta-borate (BBO) by parametric
  down conversion of light from an Ar ion pump laser (PL). After  walk-off
  compensation ($\lambda/2$, CC), down-converted light is collected behind
  interference
  filters (IF) into birefringence-compensated (FPC) single mode optical fibers
  (SMF). Polarization measurements are carried out with a
  combination of a quarter wave plate ($\lambda/4$) and polarization filters
  (PF) in front of photon counting detectors D$_{1,2}$. The measurement basis
  for each arm (1,2) is chosen by rotation of the wave plate and polarizing
  filter by angles $\alpha_{1,2}, \beta_{1,2}$ accordingly.
}
\end{figure}

\textit{Experiment.} We begin with a traditional
parametric down conversion source~\cite{kwiat:95} for
polarization-entangled photon pairs with optimized collection
geometry in single mode optical fibers~\cite{kurtsiefer:01}
(Fig.~\ref{expsetup}). Light from a continuous-wave Ar-ion laser at
351\,nm is pumping a 2\,mm thick barium-beta-borate crystal, cut for
type-II parametric down conversion to degenerate wavelengths of
702\,nm with a Gaussian spectral distribution of 5\,nm (FWHM). We
chose a pump power of about 40\,mW to ensure both single frequency
operation of the pump laser and to avoid saturation effects in the
photodetectors. Collection of down-converted light into single mode
optical fibers ensures a reasonably high polarization entanglement
to begin with. In this configuration, we observed visibilities of
polarization correlations of $>98\%$ both in the HV and
$\pm45^\circ$ linear basis for polarizing filters located before the
fibers. In order to avoid a modulation of the collection efficiency
with optical components due to wedge errors in the wave plates, we
placed subsequent polarization analyzing elements behind the fiber.

The projective polarization measurements for the different settings
of the two observers were carried out using quarter wave plates,
rotated by motorized stages by respective angles $\alpha_{1,2}$, and
absorptive polarization filters rotated by angles $\beta_{1,2}$ in a
similar way with an accuracy of 0.1~degree. This combination allows
to project on arbitrary elliptical polarization states. Finally,
photodetection was done with passively quenched silicon avalanche
diodes, and photon pairs originating from a down conversion process
were identified by coincidence detection. The compensator crystals
(CC) and fiber birefringence compensation (FPC) were adjusted such
that we were able to detect photon pairs in a singlet state.

After birefringence compensation of the optical fibers, we observed the
corresponding polarization correlations between both arms with a
visibility of $99.5\pm0.2$\% in the HV basis, $99.0\pm0.2$\% in the
$\pm45^\circ$ linear basis, and $98.2\pm0.2 $\% in the circular polarization
basis. Typical count rates were 10100\,s$^{-1}$ and 8000\,s$^{-1}$ for single
events in both arms, and about 930\,s$^{-1}$ for coincidences for orthogonal
polarizer positions. We measured an accidental coincidence rate using
a delayed detector signal of $0.41\pm0.07\,$s$^{-1}$, corresponding to a
time window of 5\,ns.

\begin{figure}
\includegraphics{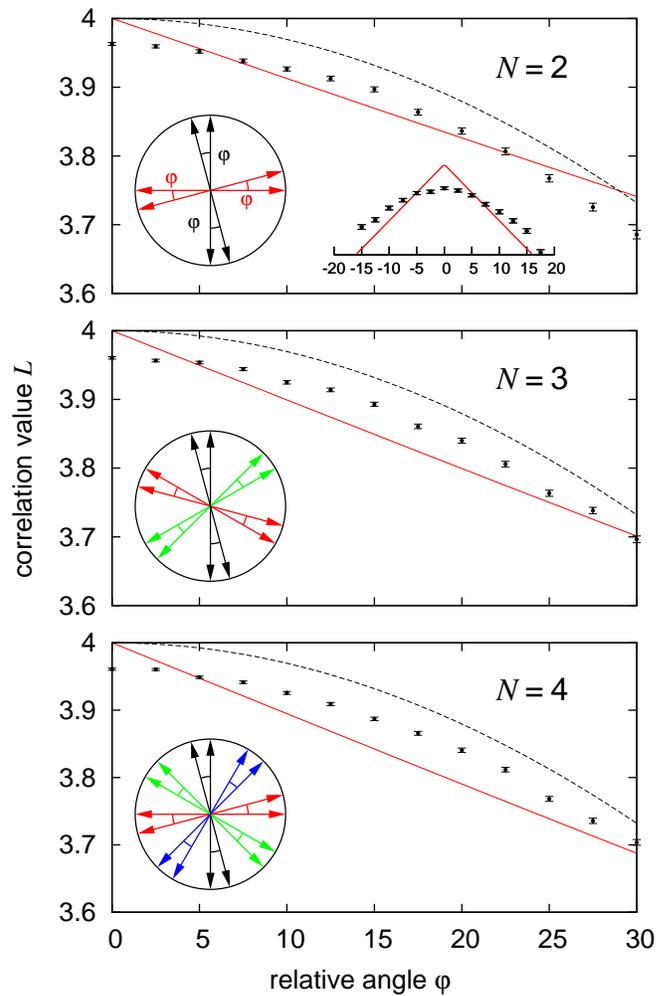}
\caption{ Experimental results for the observed
  correlation parameters $L_N$ (dots), the quantum mechanical prediction for a
  pure singlet state (curved lines, dashed), and the bounds for the non-local
  variable models (almost straight lines). In all cases, our experiment
  exceeds the NLV bounds for appropriate difference angles $\varphi$.} \label{fig:expresult}
\end{figure}

The two orthogonal planes we used in the Poincar\'e
sphere included all the linear polarizations for one,
and H/V linear and circular polarizations for the other. That
way, we intended to take advantage of the better polarization
correlations in the 'natural' basis HV for the down conversion
crystal. Each of the $4N$ correlation coefficients
$C(\vec{a},\vec{b})$ in (\ref{ineqdiscr},\ref{EjN}) was obtained
from four settings of the polarization filters via
\begin{equation}
C\big(\vec{a},\vec{b}\big)=\frac{n_{\vec{a},\vec{b}} +
  n_{-\vec{a},-\vec{b}}-n_{-\vec{a},\vec{b}} -
  n_{\vec{a},-\vec{b}}}{
  n_{\vec{a},\vec{b}}+n_{-\vec{a},-\vec{b}} +
  n_{-\vec{a},\vec{b}}+n_{\vec{a},-\vec{b}}}
\end{equation}
from the four coincident counts $n_{\pm\vec{a},\pm\vec{b}}$ obtained
for a fixed integration time of $T=4$\,s each.  For $N=2,3$ and 4, we
carried out the full generic set of 8, 12, and 16 setting groups,
respectively, with each $E_j^N(0)$ containing a HV analyzer setting.

A summary of the values of $L$ corresponding to inequalities for
$N=2$, 3 and 4 are shown in Fig.~\ref{fig:expresult}, together with
the corresponding bounds (\ref{ineqdiscr}) and the quantum
expectation for a pure singlet state (\ref{singlet}). The
corresponding standard deviations in the results were obtained
through usual error propagation assuming Poissonian counting
statistics and independent fluctuations on subsequent settings. For
$N=2$, we already observe a clear violation of the NLV bound; the
largest violation we found was for $N=4$ with
about 17 standard deviations above the NLV bound. As expected, the
experimental violation increases with growing number of averaging
settings $N$. Selected combinations of ($N,\varphi$) violating NLV
bounds are summarized in table \ref{table:summary}.

\begin{table}
\begin{tabular}{c|c|c|c|c}
$N$&$\varphi$&$L_{\rm NLV}$&$L_{\rm exp}\pm\sigma$&$L_{\rm
exp}-L_{\rm
  NLV}$\\[1mm]
\hline \hline
2&12.5$^\circ$&3.8911&$3.9127\pm0.0033$&6.45\,$\sigma$\\
2&  15$^\circ$&3.8695&$3.8970\pm0.0036$&7.59\,$\sigma$\\
2&17.5$^\circ$&3.8479&$3.8638\pm0.0042$&3.83\,$\sigma$\\
\hline
3&12.5$^\circ$&3.8743&$3.9140\pm0.0027$&14.77\,$\sigma$\\
3&  15$^\circ$&3.8493&$3.8930\pm0.0030$&14.58\,$\sigma$\\
3&17.5$^\circ$&3.8243&$3.8608\pm0.0034$&10.67\,$\sigma$\\
3&  20$^\circ$&3.7995&$3.8400\pm0.0036$&11.15\,$\sigma$\\
\hline
4&12.5$^\circ$&3.8686&$3.9091\pm0.0024$&17.01\,$\sigma$\\
4&  15$^\circ$&3.8424&$3.8870\pm0.0026$&16.84\,$\sigma$\\
4&17.5$^\circ$&3.8164&$3.8656\pm0.0029$&17.11\,$\sigma$\\
\end{tabular}
\caption{\label{table:summary}Selected values of $L$ violating the
NLV bounds $L_{\rm NLV}$ for different averaging numbers $N$.}
\end{table}

Our results are well-described assuming residual colored noise in the singlet
state preparation \cite{cabello:05}. We attribute the small asymmetry of
$L_{\rm exp}$ in $\varphi$ (see inset in Fig.~\ref{fig:expresult}) to
polarizer alignment accuracy.

\textit{Overview and Perspectives.} After the very general
motivation sketched in the introduction, we have focused on
Leggett's model. Let's now set this model in a broader picture.
Non-locality having being demonstrated, the only classical mechanism
left to explain quantum correlations is the exchange of a signal. It
is therefore natural to assume, as an alternative model to quantum
physics, that the source produces independent particles, which later
exchange some communication. Sure, this communication should travel
faster than light, so the model has to single out the frame in which
this signal propagates: it can be either a preferred frame
(``quantum ether''), in which case even signaling is not logically
contradictory \cite{ebe}; or a frame defined by the measuring
devices, in which case the model departs from the quantum
predictions when the devices are set in relative motion
\cite{sua97,zbi01}. Obviously, there are NLV models that do
reproduce exactly the quantum predictions. Explicit examples are
Bohmian Mechanics \cite{bohm} and, for the case of two qubits, the
Toner-Bacon model \cite{ton03}. Both are deterministic. Now, in
Bohmian mechanics, if the first particle to be measured is A, then
assumption (\ref{assa}) can be satisfied, but assumption
(\ref{assb}) is not. This remark sheds a clearer light on the
Leggett model, where both assumptions are enforced: the particle
that receives the communication is allowed to take this information
into account to produce non-local correlations, but it is also
required to produce outcomes that respect the marginals expected for
the local parameters alone.

As a conclusion, it must be said that the broad goal sketched in the
introduction, namely, to pinpoint the essence of quantum physics,
has not been reached yet. However, Leggett's model and its
conclusive experimental falsification reported here have added a new
piece of information towards this goal.

\textit{Acknowledgements.} We are grateful to Anthony J. Leggett,
Artur Ekert and Jean-Daniel Bancal for fruitful discussions. C.B.
acknowledges the hospitality of the National University of
Singapore. This work was partly supported by ASTAR grant
SERC-052-101-0043, by the European QAP IP-project and by the Swiss
NCCR ``Quantum Photonics".


\begin{thebibliography}{99}

\bibitem{bell} J.S. Bell, Physics {\bf 1}, 195 (1964)

\bibitem{pr94}
S.~Popescu and D.~Rohrlich, Found. Phys. \textbf{24},
379--385 (1994).

\bibitem{nlnd} See e.g. N.J.~Cerf, N.~Gisin, S.~Massar, S.~Popescu, Phys. Rev.
Lett. \textbf{94}, 220403 (2005); N. Brunner, N. Gisin, V. Scarani,
New J. Phys. {\bf 7}, 88 (2005); N. Linden, S. Popescu, A.J. Short,
A. Winter, quant-ph/0610097; H. Barnum, J. Barrett, M. Leifer, A.
Wilce, quant-ph/0611295.


\bibitem{sua97} A.Suarez, V. Scarani, Phys. Lett. A {\bf 232}, 9
(1997)

\bibitem{zbi01} H. Zbinden, J. Brendel, N. Gisin, W. Tittel, Phys. Rev. A {\bf 63}, 022111
(2001); A. Stefanov, H. Zbinden, N. Gisin, A. Suarez, Phys. Rev.
Lett. {\bf 88}, 120404 (2002)


\bibitem{leg03} A.J. Leggett, Found. Phys. \textbf{33}, 1469 (2003)

\bibitem{gro07} S. Gr\"oblacher, T. Paterek, R. Kaltenbaek, \v{C}. Brukner, M. Zukowksi, M. Aspelmeyer, A. Zeilinger, Nature \textbf{446}, 871 (2007)

\bibitem{suppl} Supplementary information of Ref.~\cite{gro07}.

\bibitem{par07} S. Parrott, arXiv:0707:3296


\bibitem{kwiat:95} P.G. Kwiat, K. Mattle, H. Weinfurter, A. Zeilinger, A.V. Sergienko, Y. Shih, Phys. Rev. Lett. \textbf{75}, 4337 (1995)


\bibitem{kurtsiefer:01} C. Kurtsiefer, M. Oberparleiter, H. Weinfurter, Phys. Rev. A \textbf{64},
023802 (2001)

\bibitem{cabello:05} A. Cabello, A. Feito, A. Lamas-Linares, Phys. Rev. A
  \textbf{72}, 052112 (2005)


\bibitem{ebe} Ph. Eberhard, A Realistic Model for Quantum Theory with a Locality
Property, in: W. Schommers (ed.), Quantum Theory and Pictures of
Reality (Springer, Berlin, 1989)


\bibitem{bohm} D. Bohm, B.J. Hiley, The undivided universe
(Routledge, New York, 1993)

\bibitem{ton03} B. F. Toner, D. Bacon, Phys. Rev. Lett. {\bf 91}, 187904 (2003)


\end{thebibliography}
\end{document}